\documentclass[twocolumn, 
prb,preprintnumbers,amsmath,amssymb,floatfix]{revtex4}

\usepackage{graphicx}
\usepackage{dcolumn}

\begin{document}

\title{Rare-Earth Tetraborides ${\cal R}$B$_4$:\\
 Analysis of Trends in the Electronic Structure}
\author{Z. P. Yin and W. E. Pickett}
\affiliation{Department of Physics, University of California Davis, 
  Davis, CA 95616}
\date{\today}
\begin{abstract}
Both the basic electronic structure of tetraborides, and the changes 
across the lanthanide series
in $R$B$_4$ ($R$ = rare earth) compounds, are studied using the correlated band theory
LDA+U method in the all-electron Full Potential Local Orbital (FPLO) code.  
A set of boron bonding bands
can be identified that are well separated from the antibonding bands. 
Separately,
the ``dimer B'' $2p_z$ orbital is non-bonding ({\it viz.} 
graphite and MgB$_2$), 
and mixes strongly with the
metal $4d$ or $5d$ states that form the conduction states. The bonding bands
are not entirely filled even for the trivalent compounds (thus the cation
$d$ bands have some filling), which accounts for the lack of stability 
of this structure when the cations are divalent (with more bonding states
unfilled).  The trends in the mean $4f$ level for both majority and
minority, and occupied and unoccupied, states are presented and interpreted. 
\end{abstract}
\maketitle

\section{Background and Motivation}
The tendency of the metalloid boron to form clusters has led to widespread
study of the properties of condensed boron.  Of the many classes of 
compounds that B forms, B-rich metal borides include classes with very
important, and intensely studied, properties.  One example is MgB$_2$,
which is the premier phonon-coupled superconductor\cite{akimitsu} 
(at 40 K). Although this structural class includes several transition
metal borides and other simple metal borides (such as LaB$_2$), MgB$_2$ is
unique in this single-member class of quasi-two-dimensional $s$-$p$ metal with 
very high superconducting transition temperature due to strong covalent
B-B bonds that are driven metallic\cite{jan} by the crystal structure and chemistry.  

Another class that has received great attention is the hexaborides MB$_6$
formed from vertex-linked B$_6$ octahedra that enclose the metal ion in
the cubic interstitial site. This class includes the divalent metals
(M=Ca, Sr, Ba) that are small gap semiconductors.\cite{lip1, lip2, Imai, Takeda, Chen, Rupp, Muranaka, Gavilano, Gavilano2, Schmitt}
The stability of
this structure was understood decades ago, when cluster studies 
established\cite{lip1,lip2} that the bonding states of 
linked B$_6$ clusters are filled
by 20 electrons, which requires two per B$_6$ unit in addition to the B valence electrons.
There are many trivalent hexaborides as well, 
including lanthanide members which
have very peculiar properties: unusual magnetic ordering, heavy fermion
formation, and superconductivity.\cite{Takeda, Chen, Gavilano2, Otani2, Chen2, Teredesai, Mo} 
Two monovalent members, 
NaB$_6$ [\onlinecite{Perkins}] and KB$_6$ [\onlinecite{Ammar}], 
have been reported.

\begin{figure}[tbp]
{\resizebox{7.8cm}{5.0cm}{\includegraphics{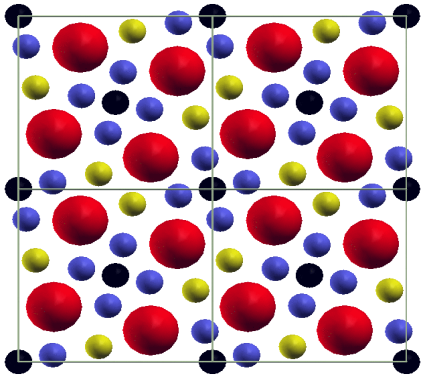}}}
\caption{(color online)
Structure of $R$B$_4$ viewed from along the c direction.
The large metal ion spheres (red) lie in $z$=0 plane.
Apical B1 atoms (small black) lie in $z$ $\simeq$ 0.2
and $z$ $\simeq$ 0.8 planes.
Lightly shaded (yellow) dimer B2 and equatorial 
B3 (dark, blue) atoms lie in $z$=0.5 plane.  The sublattice of R ions
is such that each one is a member of two differently oriented 
R$_4$ squares, and of three R$_3$ triangles.
   }
\label{structureFig}
\end{figure}

Yet another class that has been known for decades is the metal 
(mostly rare earths) tetraboride
$R$B$_4$ family, which is richer both structurally and 
electronically
and for which considerable data is available (see: for several
 $R$B$_4$, Refs. [\onlinecite{Fisk, Etourneau,lattice-all, fisk-lattice}];
 YB$_4$, Refs. [\onlinecite{Y-Lattice, Otani, Otani1, Tanaka, Gunster}];
 LaB$_4$, Ref. [\onlinecite{La-Lattice}];
 CeB$_4$, Refs. [\onlinecite{Meschel, Ce-Lattice, zalkin}];
 NdB$_4$, Ref. [\onlinecite{Nd-Lattice}];
 GdB$_4$, Refs. [\onlinecite{Mean-Gd, Ji, Gd-Lattice,Buschow, Lovesey, Blanco}];
 TbB$_4$, Refs. [\onlinecite{Mean-Tb, Gianduzzo, Tb-Lattice, Heiba, Will,Will2}];
DyB$_4$, Refs. [\onlinecite{Okuyama, Watanuki2, Matsumura, Dudnik,Schafer, Watanuki}];
 ErB$_4$, Refs. [\onlinecite{Will2,Er-and-Dy-Lattice,Derkachenko}]).
Yttrium and all the lanthanides except Eu and Pm form isostructural metallic 
tetraborides $R$B$_4$ with space group
P4/mbm (\#127), described below and pictured in Fig. \ref{structureFig}. 
Presumably Eu is not stable in the tetraboride structure because of its
preference for the divalent configuration in such compounds.  The Sr and
Ba tetraborides also are not reported.  A ``calcium tetraboride'' with
formula Ca(B$_{1-x}$C$_x$)$_4$, $x\approx 0.05$ was reported\cite{cab4}
recently.

These rare-earth tetraborides exhibit 
an unusual assortment of magnetic properties.  
While CeB$_4$ and YbB$_4$ ($f^1$ and $f^{13}$ respectively) don't order 
and PrB$_4$ orders ferromagnetically at T$_c$=25 K,\cite{Buschow} 
all of the others ($R$=Nd, Sm, Gd, Tb, Dy, Ho, Er, Tm) order 
antiferromagnetically, 
with N\'eel temperature T$_N$ (see Table I) spanning the range 7-44 K.
A noteworthy peculiarity is that T$_N$ doesn't obey de Gennes' 
scaling law, which says that 
the magnetic transition
temperature is proportional to
$(g_J-1)^2J(J+1)$ across an isostructural series where the rare-earth atom
is the only magnetic component.\cite{Will, Noakes}  (here $J$ is the Hund's
rule total angular momentum index, $g_J$ is the corresponding Land\'e
$g$-factor.)  In the rare earth nickel borocarbide series, for example,
de Gennes scaling is obeyed faithfully.\cite{deGennes}  
This lack of scaling indicates that
magnetic coupling varies across the series, rather than following a simple RKKY-like
behavior with a fixed Fermi surface.

Both the ferromagnetic member PrB$_4$ and
antiferromagnetic ones $R$B$_4$ show strong magnetic
anisotropy. For ferromagnetic PrB$_4$ the c axis is the easy axis. 
The situation is more complicated for the antiferromagnetic compounds, 
which display varying orientations  of their 
moments below T$_N$, 
and some have multiple phase transitions. GdB$_4$ and ErB$_4$ have only 
one second order phase transition, while both TbB$_4$ and DyB$_4$ 
have consecutive second order phase transitions at distinct temperatures.
A yet different behavior is shown by HoB$_4$ and TmB$_4$,
which have a second order 
phase transition followed by a first order 
phase transition at lower temperature. The magnetic ordering 
temperatures, primary spin orientations, and experimental
and theoretical effective (Curie-Weiss) magnetic moments have been collected in 
Table I.

The variety of behavior displayed by these tetraborides suggests a
sensitivity to details of the underlying electronic structure.  Unlike
the intense scrutiny that the tetraborides have attracted, there has
been no thorough study of the tetraboride electronic structure,
which contains a new structural element (the ``boron dimer") and an
apical boron that is inequivalent to the equatorial boron in the octahedron.
We provide here a detailed analysis, and in addition we provide an initial
look into the trends to be expected in the $4f$ shells of the rare
earth ions.

\begin{table}
\caption{Data on magnetic ordering in the $R$B$_4$
compounds.\cite{Fisk, lattice-all, Buschow, Watanuki}
The columns provide the experimental ordering temperature(s) T$_{mag}$, 
the ordering temperature T$_{th}$ predicted by de Gennes law (relative to
the forced agreement for the GdB$_4$ compound), the orientation of
the moments, and the measured ordered moment compared to the 
theoretical Hund's rule moment ($\mu_B$).}
\begin{tabular}{|c|c|c|c|c|c|}
\hline
        & T$_{mag}$ (K) & T$_{th}$ (K)  & direction     & $\mu$(exp) & $\mu$(th) \\
PrB$_4$ & 24            & 2.1           & $\parallel$ c & 3.20       & 3.58  \\
SmB$_4$ & 26            & 12            &    --         &   --       & 0.84  \\
GdB$_4$ & 42            & 42            & $\perp$ c     & 7.81       & 7.94   \\
TbB$_4$ & 44, 24        & 28            & $\perp$ c     & 9.64       & 9.72   \\ 
DyB$_4$ & 20.3, 12.7    & 19            & $\parallel$ c & 10.44      & 10.63  \\
HoB$_4$ & 7.1, 5.7(1st) & 12            & $\parallel$ c & 10.4       & 10.6    \\
ErB$_4$ & 15.4          &  7            & $\parallel$ c & 9.29       &9.60    \\
TmB$_4$ & 11.7, 9.7(1st) & 3            & $\perp$ c     & 7.35       & 7.56   \\
\hline
\end{tabular}
\end{table}

\section{Crystal Structure}
The full $R$B$_4$ structure was first reported by 
Zalkin and Templeton\cite{zalkin}
for the Ce, Th, and U members.
These tetraborides crystallize at
room temperature in the tetragonal space group \textit{P4/mbm, D}$_{4h}^{5}$
with four formula units
occupying the positions listed in 
Table \ref{structure}.  The lattice constants for the reported
rare earth tetraborides are presented in Table \ref{lattice}.

The B1 and B3 atoms form B$_6$ octahedra (apical and equatorial vertices,
respectively) that are connected by
B2 dimers in the $z$=1/2 plane. The B$_6$ octahedra, which are arrayed
in centered fashion in the $x$-$y$ plane within the cell, are flattened
somewhat, with distances from the center being 1.20 $\AA$ along the c axis
and 1.29 $\AA$ in the $x$-$y$ plane (taking GdB$_4$ as an example).  
Each B2
atom is bonded to two B1 atoms in separate octahedra and to one other
B2 atom.  A suggestive form for the chemical formula then is
[$R$$_2$B$_2$B$_6$]$_2$.
The rare-earth atoms lie in the large interstitial holes
in the $z$=0 plane, and form a 2D array that can be regarded as fused
squares and rhombuses.\cite{Gd-Lattice}

\begin{figure}[tbp]
\rotatebox{-90}
{\resizebox{7.8cm}{7.8cm}{\includegraphics{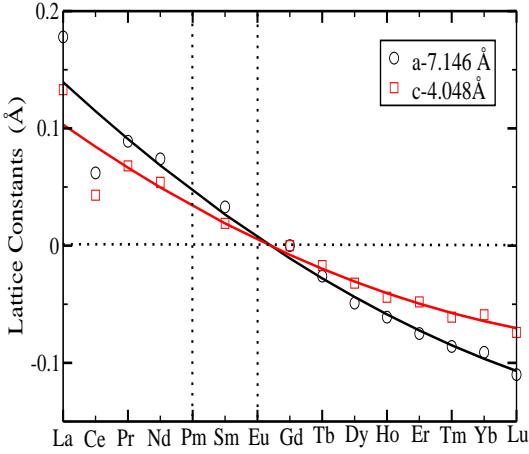}}}
\caption{
Plot of experimental lattice constants of $R$B$_4$ vs position in
the Periodic Table (atomic number), showing a lanthanide contraction
of about 5\% for $a$, 3\% for $c$.  
The smooth lines show a quadratic
fit to the data.
   }
\label{latticeFig}
\end{figure}

The $R$ site symmetry is $mm$.  The symmetry of an $R$ site
is important for the properties
of the compounds, as it dictates the crystal field splitting
of the ion with total angular momentum $\vec J = \vec L + \vec S$ and
 thereby the resulting magnetic state at low temperature.
The $R$ ion is coordinated by seven B atoms in planes both above and below,
three of them being dimer B2 atoms (two 2.88~\AA~distant and one at a
distance of 3.08~\AA) and four of them equatorial B3 atoms (two each at
distances of 2.76~\AA~ and 2.84~\AA). 
Within the unit cell the four R sites
form a square of side $d=0.518a$ = 3.70\AA, oriented at 
about 15$^{\circ}$ with respect to
the square sublattice of B$_6$ octahedra.
The (low) site symmetries of the apical B1, dimer B2, and equatorial B3 
atoms are $4, mm, m,$ respectively.

The reported lattice constants for the lanthanides are plotted in Fig.
\ref{latticeFig}.  It is evident that most fall on smooth lines reflecting
the lanthanide contraction in this system.  The behavior is representative
of trivalent behavior, from La through to Lu.  The big exception is
Ce, which has smaller volume suggesting that, rather than being simple 
trivalent, the $4f$ electron is participating in bonding.  Pm with all
unstable isotopes has not been reported.  EuB$_4$ also has not been reported; Eu
typically prefers the divalent state (due to the gain in energy of the
half-filled $4f$ shell) so it is not surprising that it is different.
However, some divalent tetraborides do form in this structure 
({\it e.g.} CaB$_4$, see Sec. IV) so it cannot
be concluded that EuB$_4$ is unstable simply on the basis of
divalency.  Finally, the small deviation of Yb from the smooth curves
suggest it maybe be mixed or intermediate valent (although close 
to trivalent).

\begin{table}
\caption{Site designations, symmetries, and atomic positions of the atoms in the
$R$B$_4$ crystal.}
\label{structure}
\begin{tabular}{cccc}
\hline
\hline
$R$  &4g  & mm       & (x, $\frac{1}{2}$+x, 0)\\

B1   &4e  & 4       & (0, 0, z)\\

B2   &4h  & mm       & (x, $\frac{1}{2}$+x, $\frac{1}{2}$)\\

B3   &8j  & m       & (x, y, $\frac{1}{2}$)\\
\hline
\hline
\end{tabular}
\end{table}

\begin{table}[th]
\caption{Tabulation of the lattice constants
and internal structural
parameters used in our calculations. Considering the extreme regularity of the 
internal coordinates through this system, the irregularity in $z_{B1}$
for Dy should be 
treated with skepticism.}
\label{lattice}
\begin{centering}
\begin{tabular}{|c|c|c|c|c|c|c|c|c|}
\hline
 R & a($\AA$)   & c($\AA$)   &$x_R$       & $z_{B1}$   & $x_{B2}$   & $x_{B3}$   & $y_{B3}$& Ref. \\
\hline
 Y & 7.111      & 4.017      & 0.318       & 0.203     & 0.087       & 0.176     & 0.039 &\onlinecite{Y-Lattice} \\
\hline
La & 7.324      & 4.181      & 0.317       & 0.209     & 0.088       & 0.174     & 0.039 &\onlinecite{fisk-lattice},\onlinecite{La-Lattice} \\
\hline
Ce & 7.208      & 4.091      & 0.318       & 0.203      & 0.087      & 0.176      & 0.039&\onlinecite{fisk-lattice},\onlinecite{Ce-Lattice} \\
\hline
Pr & 7.235      & 4.116      & 0.318       & 0.203      & 0.087      & 0.176      & 0.039&\onlinecite{lattice-all} \\
\hline
Nd & 7.220      & 4.102      & 0.318       & 0.203      & 0.087      & 0.176      & 0.039&\onlinecite{fisk-lattice},\onlinecite{Nd-Lattice} \\
\hline
Pm & 7.193      & 4.082      & 0.318       & 0.203      & 0.087      & 0.176      & 0.039&   \\
\hline
Sm & 7.179      & 4.067      & 0.318       & 0.203      & 0.087      & 0.176      & 0.039&\onlinecite{lattice-all} \\
\hline
Eu & 7.162      & 4.057      & 0.318       & 0.203      & 0.087      & 0.176      & 0.039&    \\
\hline
Gd & 7.146      & 4.048      & 0.317       & 0.203      & 0.087      & 0.176      & 0.038&\onlinecite{Gd-Lattice} \\
\hline
Tb & 7.120      & 4.042      & 0.317       & 0.202      & 0.087      & 0.176      & 0.039&\onlinecite{Tb-Lattice},\onlinecite{Will} \\
\hline
Dy & 7.097      & 4.016      & 0.319       & 0.196      & 0.086      & 0.175      & 0.039&\onlinecite{lattice-all},\onlinecite{Er-and-Dy-Lattice} \\
\hline
Ho & 7.085      & 4.004      & 0.318       & 0.203      & 0.087      & 0.176      & 0.039&\onlinecite{lattice-all} \\
\hline
Er & 7.071      & 4.000     & 0.318       & 0.203      & 0.086       & 0.177     & 0.038&\onlinecite{Will},\onlinecite{Er-and-Dy-Lattice} \\
\hline
Tm & 7.057      & 3.987     & 0.318       & 0.203      & 0.087      & 0.176      & 0.039&\onlinecite{fisk-lattice} \\
\hline
Yb & 7.064      & 3.989     & 0.318       & 0.203      & 0.087      & 0.176      & 0.039&\onlinecite{fisk-lattice} \\
\hline
Lu & 7.036      & 3.974     & 0.318       & 0.203      & 0.087      & 0.176      & 0.039&\onlinecite{fisk-lattice} \\
\hline
\end{tabular}
\end{centering}
\end{table}

\section{Calculational Methods}
The full potential local orbital (FPLO) code\cite{FPLO} (version 5.18) was
used in our calculations. Both LDA (PW92 of Perdew and Wang\cite{PW92}) 
and LDA+U (using the atomic limit functional) are used. 
We used a k mesh of 12$^3$  in the full Brillouin zone.  For the
density of states (DOS) plot and
Fermi surface plot, we used a k mesh of 24$^3$ for more precision.
The basis set was  1s2s2p3s3p3d4s4p::(4d4f5s5p)/6s6p5d+ for 
all metal elements(except Y(1s2s2p3s3p3d::(4s4p)/5s5p4d+) 
and Ca(1s2s2p::(3s3p)/4s4p3d+)).
For boron atoms we used the basis ::1s/(2s2p3d)+. 

In the LDA+U calculations we used values typical for $4f$ atoms U = 8 eV and J = 1 eV
(corresponding to Slater integrals F$_1$=8.00, F$_2$=11.83, F$_4$=8.14, F$_6$=5.86) 
throughout all calculations.  The high symmetry points in the tetragonal zone
are $\Gamma$=(0,0,0), 
$X=(\frac{\pi}{a},0,0)$, 
$M=(\frac{\pi}{a},\frac{\pi}{a},0)$, 
$Z=(0,0,\frac{\pi}{c})$,
$R=(\frac{\pi}{a},0,\frac{\pi}{c})$, and
$A=(\frac{\pi}{a},\frac{\pi}{a},\frac{\pi}{c})$.

\section{General Electronic Structure}
\begin{figure}[tbp]
\rotatebox{-90}
{\resizebox{6.4cm}{9.3cm}{\includegraphics{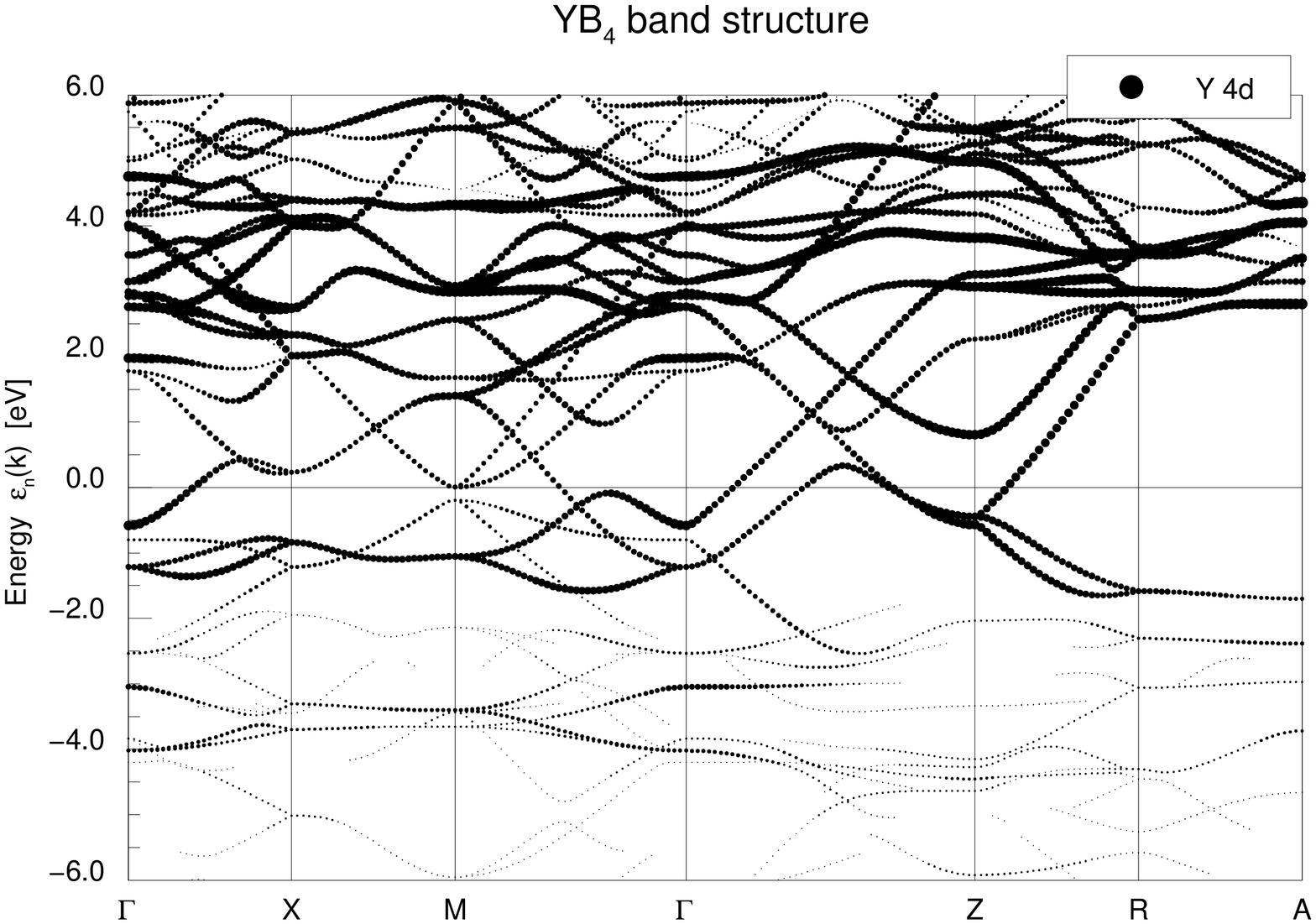}}}
{\rotatebox{-90}
{\resizebox{6.4cm}{9.3cm}{\includegraphics{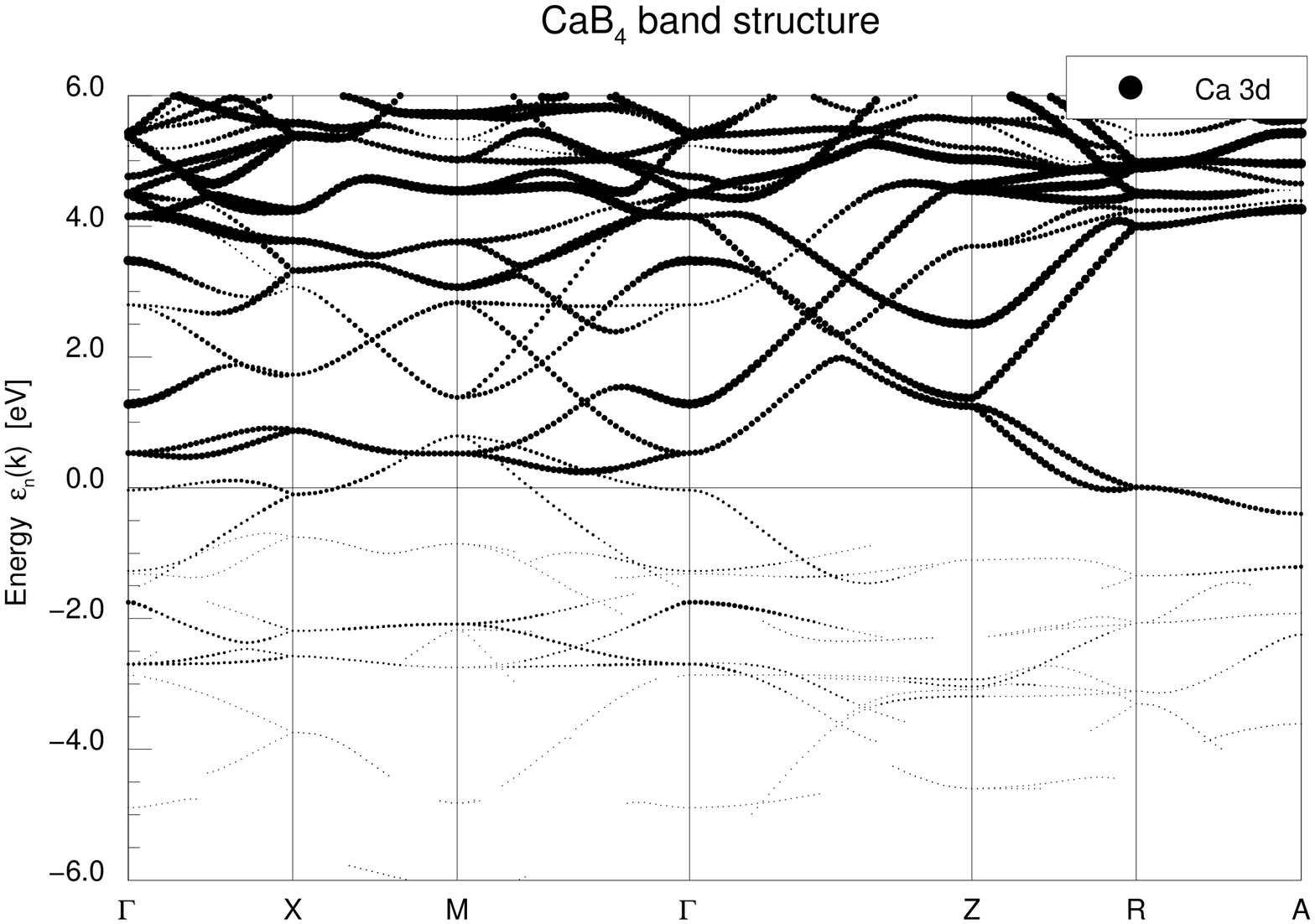}}}}
\vskip0.5in
\caption{
Band structure of YB$_4$ (top panel) and CaB$_4$ (lower panel) 
within 6 eV of the Fermi level
along high symmetry directions, showing
the gap that opens up around E$_F$ (taken as the zero of energy)
throughout much of the top and bottom portions of the tetragonal Brillouin
zone.  Notice the lack of dispersion along the upper and lower zone
edges R-A-R ($k_z$=$\pi/c$, and either $k_x$ or $k_y$ is $\pi/a$).
Note also that, due to the non-symmorphic space group, bands stick
together in pairs along X-M (the zone `equator') and along R-A (top and bottom
zone edges).}
\label{YB4_band}
\end{figure}

The valence-conduction band structure of YB$_4$ (where there are
no $4f$ bands) is shown in Fig. \ref{YB4_band}.  For LaB$_4$, which differs 
in volume and conduction $d$ level position, the bands
are very similar, with only slightly differing Fermi level crossings
along the M-$\Gamma$ direction.  The occupied valence
bandwidth is 11 eV (not all bands are shown in this figure).  One striking 
feature of the bands is the broad gap of more than 3 eV along the
top (and bottom) edges R-A-R of the Brillouin zone.  
Bands along these lines stick together in pairs due to the non-symmorphic
space group, and nearly
all bands disperse very weakly with $k_x$ (or $k_y$) along these
edges.  This gap closes
along the $k_z = \pi/c$ plane of the zone only for small in-plane
components of the wavevectors.  It is such gaps enclosing
E$_F$ that often account for the stability of a crystal structure,
and the stability of boride structures, including this one,
has been a topic of interest for 
decades.\cite{lip1,lip2,lundstrom,vegas}

The band structure of a divalent cation member (CaB$_4$) is also
included in Fig. \ref{YB4_band} for comparison.  The largest difference
is the band filling, as expected, although some band positions differ in
important ways near the Fermi level.  Still the $3d$ bands of Ca are not
quite empty, as a band with substantial $3d$ character lies at E$_F$
at R and is below E$_F$ all along the R-A line.
CaB$_4$ can be fairly
characterized, though, as having nearly filled bonding B $2p$ bands
and nearly empty Ca $3d$ bands.

\begin{figure}[tbp]
\rotatebox{-90}
{\resizebox{7.8cm}{7.8cm}
   {\includegraphics{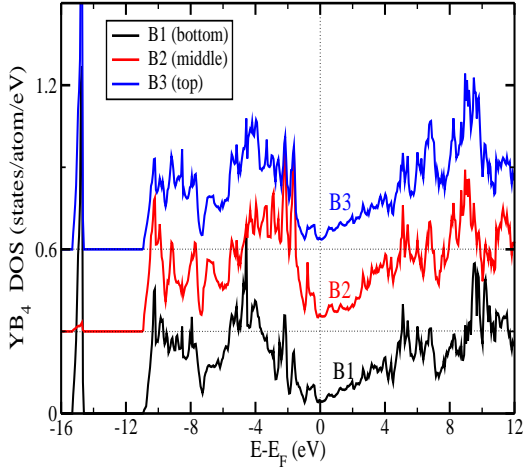}}}
\caption{
Projected density of states per atom of each of the B atoms
for YB$_4$. The curves are shifted to enable easier identification
of the differences.  The B $2p$ bonding-antibonding gap can be identified
as roughly from -1 eV to 4-5 eV.
   }
\label{B_total_dos}
\end{figure}

\begin{figure}[tbp]
\rotatebox{-90}
{\resizebox{7.8cm}{7.8cm}{\includegraphics{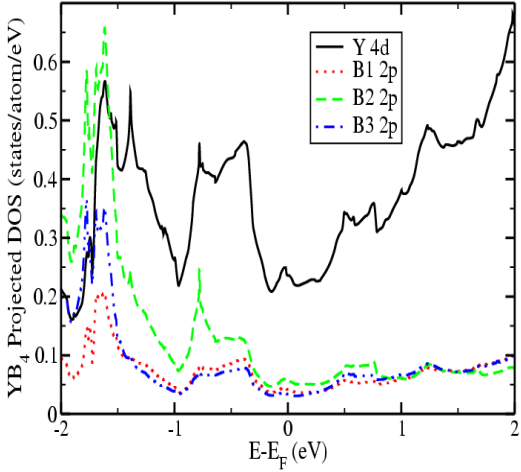}}}\caption{
Enlargement of the partial densities of states of
Y $4d$ and B $2p$ states (per atom) near the Fermi level.  The states at
the Fermi level, and even for almost 2 eV below, have strong $4d$ character.
The apical B2 character is considerably larger than that of B1 or B3 in the
two peaks below E$_F$, but is only marginally larger exactly at E$_F$.
   }
\label{YB4_PDOS}
\end{figure}

\subsection{Bonding and Antibonding Bands}
As mentioned in the Introduction, the stability of the hexaborides
is understood in terms of ten bonding molecular orbitals of the B$_6$
octahedron.  This octahedron occurs also in these tetraborides, along
with one additional B$_2$ dimer that is bonded only in the layer ($sp^2$).
Lipscomb and Britton\cite{lip1,lip2} started from this point, and
argued that each of the B2 atoms in a dimer forms single bonds
with two B3 atoms but a double bond with its dimer neighbor, so
each B2 atom needs four electrons.  The total of 20+8 electrons for
each set of 6+2 boron atoms leaves a deficit of four electrons, or a
deficit of 8 electrons in the cell.  This amount of charge can be
supplied by four divalent cations, with CaB$_4$ as an
example.  Most tetraborides contain trivalent cations, however, so this
is an issue worth analyzing.

An empirical extended H\"uckel band structure study\cite{cab4} 
for CaB$_4$ indeed
gave a gap, albeit a very narrow one.  The H\"uckel method
can be very instructive but is not as accurate
as self-consistent density functional methods.
Our FPLO calculation on CaB$_4$, shown in Fig. 3, gives a metallic band structure.  
However, the `valence' (occupied) and `conduction' (unoccupied) bands
in the bands (H\"uckel, and also FPLO) are readily identified, and it clear that
there are disjoint sets of bands with different characters.  There are the
boron bonding bands (at E$_F$ and below) that can be clearly distinguished
from conduction bands at and above E$_F$.  These conduction bands are
primarily metal $d$ bands (with an interspersed nonbonding B2 $p_z$ band,
see below). If they were $\sim$0.5 eV higher it would result in an
insulating band structure in CaB$_4$.  The boron antibonding bands lie
higher, above 5 eV at least and mix strongly with the metal $d$ bands.

The separation into bonding and antibonding B $2p$ bands agrees (almost) with the 
ideas of Lipscomb and Britton, and confirms their counting arguments.
However, the existence of numerous $R$$^{3+}$B$_4$ compounds and only
one divalent member shows
that the extra electron is not a destabilizing influence, while
it increases the conduction electron density (hence, the conductivity,
and magnetic coupling). 

In covalently bonded materials it is common to be able to identify the
distinction 
between the bonding bands and the antibonding bands.  In covalent
semiconductors, for example,
they lie respectively below and above the band gap, an absolutely clean
separation.  In the $R$B$_4$ system the $d$ bands lie within the
corresponding bonding-antibonding gap and complicate the picture.  Analysis
of the orbital-projected bands clarify this aspect.  The B1 and B3 atoms,
being engaged in three-dimensional bonding (within an octahedron {\it and}
to another unit [octahedron or dimer]), have a clear 
bonding-antibonding splitting of a few
eV (beginning just below E$_F$).  Likewise, the dimer B2 $p_x, p_y$ states
display a similar splitting.

The B2 $p_z$ orbital is quite different. As is the case in MgB$_2$ (whose
planar structure is similar to the local arrangement of a B2 atom), 
$p_z$ bands extend continuously through the gap in the B 
bonding/antibonding bands,
and mix fairly strongly with the rare earth $d$ states in that region.
There is considerable B2 $p_z$ character in the bands near (both below and above)
E$_F$ at the zone edge $M$ point, as well as the Y $4d$ character that is
evident in Fig. \ref{YB4_band}.  So while there is some B1 and B3 character
in the rare earth metal $d$ bands that lie within the boron bonding-antibonding
gap, the amount of B2 $p_z$ character is the primary type of B participation
in these bands that provide conduction and magnetic coupling.

\subsection{Pseudogap in the Density of States}
From the projected DOS of the
three types of B atoms of YB$_4$ (see
Fig. \ref{B_total_dos}), one can detect only relatively
small differences in the distribution of 
B1, B2, and B3 character arising from
their differing environments. First, note that in the DOS of
B1 and B3 there is a peak around -15 eV, while there is no such peak
for B2. This peak arises from the overlap of $2s$ and $2p_{\sigma}$
states of each of the boron atoms
forming the B$_6$ octahedra (B1 and B3); the $2s$ character is about
three times as large as the $2p_{\sigma}$ character, and
the remaining $2s$ character is mixed into the lower $2p$ bands. 
This state is a well localized B$_6$ cluster orbital, and 
there are two such orbitals (octahedral clusters) per cell.
The bridging B2 atoms
do not participate in any such bound state.

Another difference in characters of the B sites is that, in the 
region below but within 2 eV of the Fermi level, the DOS
of the dimer B2 atom is significantly larger than that of 
B1 and B3 atoms, as can be
seen in Fig. \ref{YB4_PDOS}. Together with plots showing the
band character (not shown), 
this difference reflects the fact that all of the $2p$
orbitals of B1 and B3 (octahedron) atoms are incorporated into bonding (filled)
and antibonding (empty) bands.
The distinct characteristic of the B2 $p_z$ state was discussed in
the previous subsection.
All B $2p$ states do hybridize to some degree with the metal $d$ bands, however,
and all B atoms have some contribution at the Fermi level.

\begin{figure}[tbp]
{\resizebox{6.8cm}{6.8cm}{\includegraphics{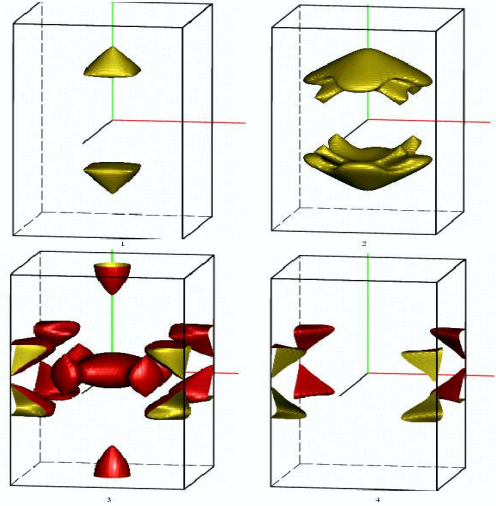}}}
\caption{ (color online)
Fermi surfaces of YB$_4$.  Light (yellow) surfaces enclose holes, dark
(red) surfaces enclose electrons.  The wide gap the throughout the
top and bottom edges of the zone account for the lack of Fermi
surfaces there. 
   }
\label{YB4FS}
\end{figure}

The full Y $4d$ DOS (not shown) establishes that these bands are centered
about 4 eV above E$_F$, with a `bandwidth' (full width at half maximum) 
of 6-7 eV (a `full bandwidth' would be somewhat larger).
The largest Y character near E$_F$ along symmetry lines is $4d(x^2-y^2)$,
primarily in the bands dispersing up from -0.5 eV at $Z$ toward $\Gamma$. 
The flat bands around -1 eV along $\Gamma-X-M-\Gamma$ are strongly $4d(z^2)$
character, indicative of a nonbonding, almost localized state in the $x$-$y$
plane.  Note that these bands disperse strongly upward along ($0,0,k_z)$
and lie 3-4 eV above E$_F$ in the $k_z=\pi/c$ plane.  Thus the $4d(z^2$) orbitals
form two nearly separate one-dimensional bands along $k_z$, and give rise to
flat parts of some Fermi surfaces (see following subsection).  These bands
can be modeled by a tight-binding band $-t_{dd\sigma}$cos$k_zc$
with hopping amplitude 
$t_{dd\sigma} \approx 1$ eV.  Most of the
$4d(xz), 4d(yz)$ character and $4d(xy)$ character lies above E$_F$, and is
centered 3-4 eV above E$_F$.  
The B2 $2p_z$ state mixes primarily with Y $4d_{xz}$, $4d_{yz}$ near the 
$M$ point (near E$_F$ and above).
The B2 $2p_z$ orbitals are shifted up somewhat with respect to the
$2p_x$, $2p_y$ states by the ligand field effects (there is a
bonding interaction within
the $x$-$y$ plane only).

\subsection{Fermi Surface}
The Fermi surfaces of YB$_4$, shown in Fig. \ref{YB4FS}, will be representative
of those of the trivalent $R$B$_4$ compounds although small differences may
occur due to element-specific chemistry of trivalent rare earths and due to
the lanthanide contraction.  The large gap along the R-A-R edges precludes
any FS on or near most of the $k_z =\frac{\pi}{c}$ face.  
The Fermi surfaces can be
pictured as follows.   Square hole pyramids with only 
slightly rounded vertices lie midway along
the $\Gamma-Z$ line, and similar nested electron pyramids lie along the $M-A$ line
near the $M$ point.  A pointed ellipsoid oriented along $k_z$ sits at the
$Z$ point.  Surrounding $\Gamma$ is lens-type electron surface joined to pointed
ellipsoids along the (110) directions.  Finally, there are two ``tortoise shell''
shaped hole surfaces within the zone, centered along the $\Gamma-Z$ lines.   

These surfaces, and the small variation through the lanthanide series,
is surely relevant to the varying magnetic behavior observed in $R$B$_4$
compounds.  There are nesting possibilities between the bases of the square
pyramids, for example, which will appear as RKKY coupling as the 
associated nesting vectors.  The ellipsoidal attachments on the zone-centered
lens surface may provide some weak nesting.

\begin{figure}[tbp]
\rotatebox{-90}
{\resizebox{8.4cm}{8.4cm}{\includegraphics{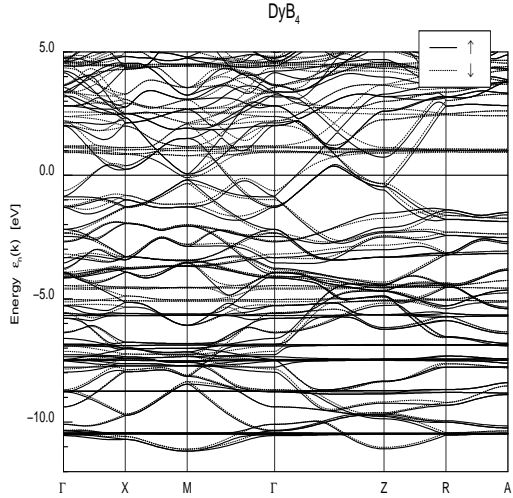}}}
\caption{
The full valence band structure of DyB$_4$, and up to 5 eV in the
conduction bands.  This plot is for ferromagnetic alignment of the
spin moments, with the solid bands being majority and the lighter, dashed
lines showing the minority bands.
The flat bands in the -4.5 eV to -11 eV are $4f$
eigenvalues as described by the LDA+U method.  
   }
\label{Dybands}
\end{figure}

\begin{figure}[tbp]
\rotatebox{-90}
{\resizebox{8.4cm}{8.4cm}{\includegraphics{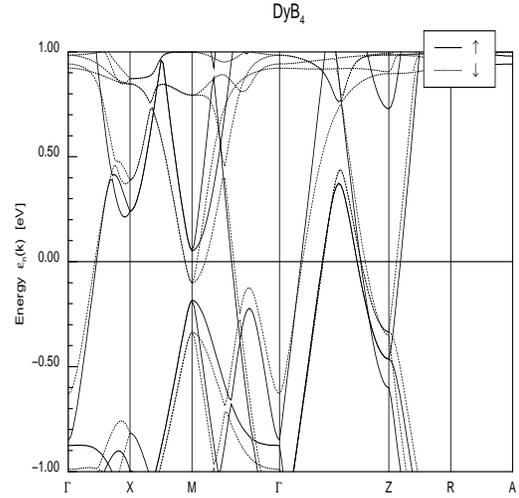}}}
\caption{
Band structure of DyB$_4$ on a fine scale around the Fermi energy,
see Fig. \ref{Dybands}.  The exchange splitting (between solid and
dashed bands) gives a direct measure of the coupling between the
polarized Dy ion and the itinerant bands (see text).  
   }
\label{Dyband2}
\end{figure}

\begin{figure}[tbp]
{\resizebox{7.8cm}{7.8cm}{\includegraphics{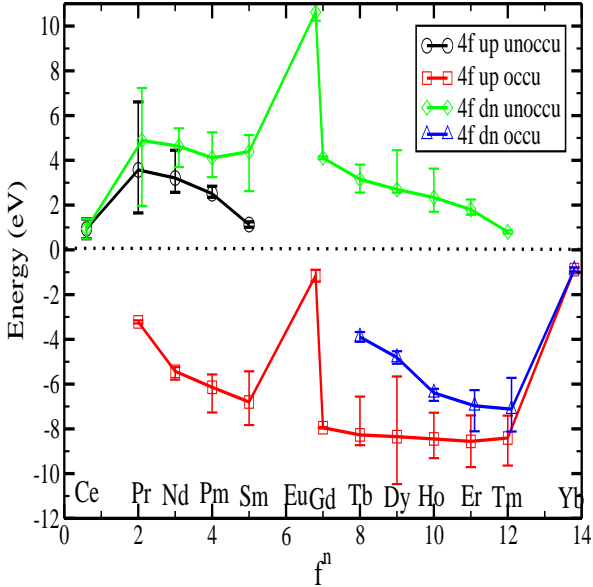}}}
\caption{
Calculated mean $4f$ eigenvalue position (symbols connected by
lines) with respect to E$_F$, and the spread in eigenvalues,
of $R$B$_4$ compounds.  The smooth
behavior from Pr to Tm (except for Eu) reflects the common trivalent
state of these ions.  Eu and Yb are calculated to be
divalent and deviate strongly
from the trivalent trend. Ce has a higher
valence than three, accounting for its deviation from the trivalent trend.
   }
\label{4flevels}
\end{figure}

\section{The Lanthanide Series}
Any effective one-electron treatment of the electronic structure
of $4f$ electron systems faces serious challenges.
The root of the difficulty is that the ground state of an open $4f$
shell has intrinsic many-body character, being characterized by the
spin $S$ and angular momentum $L$ of all of the $4f$ electrons, and
the resulting total angular momentum $J$, following Hund's rules. 
Although it is possible to delve into the extent to which the LDA+U
method can reproduce the $z$-components of such configurations,\cite{MDJ}
that is not the intention here.  LDA+U reliably gets the high spin
aspect, which contains much of the physics that determines relative
$4f$ level positions and hence trends across the series.  There is
recent evidence from calculations on rare earth nitrides\cite{larson}
that, if spin-orbit coupling is neglected and the
symmetry is lowered appropriately, the high orbital moment (Hund's
second rule) can usually be reproduced.  The exceptions are the usual
difficult (and interesting) cases of Eu and Yb.

The Hund's rule ground state of the ion often breaks the local symmetry
of the site, and if one is exploring that aspect the site symmetry should be
allowed to be lower than the crystalline symmetry.  As stated, we are
not interested here in those details.  In the calculations reported
here, the crystal symmetry is retained.  The site symmetry of the
lanthanide ion is already low ($mm$), reflected in its 14-fold
coordination with B atoms.  In addition, spin-orbit coupling has not
been included.

\subsection{Band structure}
Most of the $R$B$_4$ lanthanide tetraborides follow the usual trivalent
nature of these ions, and the itinerant parts of their band structures
are very similar to those of YB$_4$ and LaB$_4$.  The exceptions are
$R$ = Eu and Yb, which tend to be divalent to achieve a half-filled
or filled shell, respectively.  

By way of illustration of the complexity of the full $R$B$_4$ bands,
the full band structure of DyB$_4$ is presented in Fig. \ref{Dybands} 
for ferromagnetic ordering.  The $4f$ bands themselves can be identified by
their flat (weakly hybridizing) nature.  An enlarged picture of the bands
within 1 eV of E$_F$ is given in Fig. 8.  The splitting of the majority
and minority itinerant bands provide a direct measure of the Kondo
coupling of the $4f$ moment to the band states. Note that the sign of
this splitting can vary from band to band.

Figure 8 suggests that the Fermi surfaces will be different in the
magnetic tetraborides (compared to YB$_4$) in specific ways.  For Dy, the 
$\Gamma$-centered surface splits almost imperceptibly.  The surfaces that 
cross the $\Gamma$-Z line also are relatively unaffected by
exchange splitting.  At the 
M point, however, a new surface appears due to the magnetism: an
electron surface of minority spin.  For this band, the polarization
is opposite to the direction of the Dy spins.  This figure is specifically for
ferromagnetic alignment, while DyB$_4$ actually orders antiferromagnetically
(see Sec. I).  

\subsection{Position of $4f$ Levels}
The mean position of $4f$ levels is displayed in Fig. \ref{4flevels},
separated into occupied and unoccupied, and majority and minority, and
trends are more meaningful than absolute energies.
Simple ferromagnetic alignment is used here, in order to follow the 
chemical trends in the simplest manner.  
For the occupied majority states, the $4f$ level drops 
rapidly from Pr (-3 eV) to Sm (-7 eV), then becomes almost flat
for Gd-Tm (around -8 eV).  For the unoccupied minority states, the mean
$4f$ level drops almost linearly from Pr (+5 eV) to Er (+2 eV), and 
for Tm the $4f$ level is very close to E$_F$.  The unoccupied majority
levels, which become occupied minority levels beyond the middle of the
series, drop more steeply, with slope almost -1 eV per unit increase in
nuclear charge.

There are the usual exceptions to these overall trends.  Ce is very  
different, indicating that it is very untypical (the calculational result
is tetravalent and nonmagnetic).  Both Eu and Yb are divalent in the
calculation; an `extra' $4f$ state is occupied so their mean $4f$ level
position is 6 eV (8 eV for Yb) higher than the trivalent line.  

The spread in $4f$ eigenvalues is also displayed in Fig. \ref{4flevels}.
This spread is sensitive to the specific configuration that is obtained,
and also has no direct relation to spectroscopic data, although it
does reflect some of the internal potential shifts occurring in the
LDA+U method.  The distinctive features are unusually large spread for
the occupied majority levels in Dy (two electrons past half-filled
shell), and for the unoccupied minority (and also unoccupied majority)
levels in Pr (two electrons
above the empty shell).

\section{Summary}
In this paper we have provided an analysis of the electronic structure
of trivalent tetraborides, using YB$_4$ as the prototype, and compared
this with a divalent member CaB$_4$.  In agreement
with earlier observations on the likely bonding orbitals in the B atoms,
it is found that bonding states are (nearly) filled and antibonding
states are empty.  The states at the Fermi level in the trivalent
compounds are a combination of the (dimer) B2 $p_z$ nonbonding orbitals
whose bands pass through the bonding-antibonding gap, and the cation
$d$ orbitals.  Since the extra electron in the trivalent compounds does
not go into an antibonding state, there is no significant destabilization
of the crystal structure.

The trends in the energy positions of the $4f$ states in the rare earth
tetraborides has been found to be consistent with expectations based on
other rare earth systems, as is the fact that Eu and Yb tend to be
divalent rather than trivalent.  Investigations of the magnetic 
behavior of rare earth tetraborides will require individual study.
Nearest neighbor magnetic interactions may involve a
combination of $4f-4d-2p_z-4d-4f$ interactions, and longer range
RKKY interactions that may bring in the Fermi surface geometry.  Another
possible coupling path is the direct $4f-2p_z-4f$ path. The coupling
is likely to be even more complicated than in the rocksalt EuO and Eu
chalcogenides, where competition between direct and indirect magnetic
coupling paths has received recent attention.\cite{kunes}  The tetraboride
structure is fascinating in several respects.  A relevant one, if
coupling does proceed directly via $4f-2p_z-4f$, is that
the (dimer) B2 atom coordinates with {\it three} neighboring rare earths
ions, which will introduce frustration when the interaction has
antiferromagnetic sign.  

\section{Acknowledgments}
We have benefited from discussion of the calculations with
D. Kasinathan, K. Koepernik, and M. Richter, and from communication
about data on DyB$_4$ with E. Choi. 
Support from the Alexander von Humboldt Foundation,
and the hospitality of IFW Dresden, during the early part of this work
is gratefully acknowledged.  This work was supported by National Science
Foundation Grant No. DMR-0421810.

\end{document}